\documentclass[letterpaper,11pt]{article}

 \title{Value at risk and the diversification dogma}
 \author{Arturo Erdely\thanks{Personal website https://sites.google.com/site/arturoerdely}}
 \date{\small{Facultad de Estudios Superiores Acatl\'an \\
              Universidad Nacional Aut\'onoma de M\'exico \\
							\texttt{arturo.erdely@comunidad.unam.mx}\\}}
 
 \usepackage{amsthm}
 \usepackage{amssymb}
 \usepackage{amsmath}
 \usepackage{enumerate}
 \usepackage{graphicx}

\newcommand{\prob}{\mathbb{P}}
\newcommand{\esper}{\mathbb{E}}
\newcommand{\med}{\mathbb{M}}
\newcommand{\vari}{\mathbb{V}}

\theoremstyle{plain}
\newtheorem{teo}{Theorem}[section]
\newtheorem{lema}{Lemma}[section]
\newtheorem{cor}{Corollary}[section]
\newtheorem{prop}{Proposition}[section]

\theoremstyle{definition}
\newtheorem{defn}{Definition}[section]
\newtheorem{ejem}{Example}[section]
  
\theoremstyle{remark}

\begin{document}
 
\maketitle

\begin{abstract}
  \noindent The so-called risk diversification principle is analyzed, showing that its convenience depends on individual characteristics of the risks involved and the dependence relationship among them.
\end{abstract}


\section{Introduction}

A popular proverb states \textit{don't put all your eggs in one basket} and it is implicitly based on a \textit{principle} (let's call it that way momentarily) of \textit{risk diversification} which could have the following ``justification'': Suppose it is needed to take $2n$ eggs from point A to point B, walking distance, and that there are only two alternatives available, either one person carrying all the eggs in one basket, or two people with $n$ eggs each in separate (and independent) baskets. The proverb suggests that there is a higher risk with the single person alternative since if he/she happens to stumble and fall we would have a total loss, while with the second alternative only half of the eggs would be lost, and in a worst case scenario (with lower probability) where the two people fall the loss would be the same as in the first alternative, anyway.

\medskip

Let $X$ be a random variable which counts how many eggs are lost under the first alternative (one basket), and let $Y$ account for the same but for the second alternative (two baskets). Let $0<\theta<1$ be the probability of falling and breaking the eggs in a basket while walking from point A to B. Then $X$ and $Y$ are discrete random variables such that $\prob(X\in\{0, 2n\})=1$ and $\prob(Y\in\{0,n,2n\})=1,$ with point probabilities $\prob(X = 2n)=\theta,$ $\prob(X = 0) = 1-\theta,$ $\prob(Y=2n)=\theta^{\hspace{0.3mm}2},$ $\prob(Y=n)=2\theta(1-\theta),$ and $\prob(Y=0)=(1-\theta)^2.$ Certainly the probability of facing the maximum loss of $2n$ eggs has a higher probability under the first alternative, but it is also true that the \textit{no loss} probability is also higher under such alternative. Moreover: $$\prob(Y>0)\,=\,\theta^{\hspace{0.3mm}2}+2\theta(1-\theta)\,=\,\theta(2-\theta)\,>\,\theta\,=\,\prob(X>0)\,,$$ which means that there is a higher probability of suffering a (partial or total) loss under the second alternative. Therefore... does it mean that it is better to put all the eggs in one basket? If a single trip is going to take place, the answer would be yes, but if the same trip is going to be repeated a large number of times we should analyze the \textit{long run} average loss, which would be $\esper(X)=2n\theta$ for the first alternative, and $\esper(Y)=2n\theta^{\hspace{0.3mm}2}+2n\theta(1-\theta)=2n\theta$ for the second alternative, that is, in the long run there is no difference between the two alternatives.

\medskip

Is it never more convenient to diversify in two baskets? If the probability of stumbling and falling with $2n$ eggs is the same as with half of them (which might be true up to certain value of $n$) then the proverb is certainly wrong, but maybe for a sufficiently large value of $n$ we should consider different probabilities of falling and breaking the eggs, say $\theta_1$ for the first alternative and $\theta_2$ for the second one, with $\theta_1>\theta_2.$ This last condition leads to $\esper(X)>\esper(Y)$ and in such case it is more convenient to diversify if a large number of trips are going to be made. But for a single trip decision the condition $\theta_1>\theta_2$ is not enough to prefer diversification unless $\theta_2(2-\theta_2)<\theta_1\,,$ since $\theta_2<\theta_2(2-\theta_2).$

\medskip

The main purpose of the present work is to show that the common belief that risk diversification is \textbf{always} better is more a \textbf{dogma}\footnote{A system of principles or tenets; doctrine. A specific principle of a doctrine put forth, such as by a church. \textit{Source:} WordReference Random House Learner's Dictionary of American English \textcopyright \,2016.} rather than a general principle that has been proved, and that the correct view is to state that risk diversification may be better, as good as, or worse than lack thereof, depending on the risks involved and the dependence relationship among them.

\section{Risk measures}

Let $X$ be a continuous random variable, with strictly increasing distribution function $F_X,$ that represents an economic loss generated by certain events covered by insurance or related to investments. Without loss of generality we consider amounts of constant value over time (inflation indexed, for example). As a point estimation for a potential loss we may use the mean or the median. In the present work the median is preferred since it always exists for continuous random variables and it is robust, in contrast with the mean that may not exist or could be numerically unstable under heavy-tailed probability distributions. Using the quantile function (inverse of $F_X$) we calculate the median as $\med(X)=F_X^{\,-1}(\frac{1}{2})$ since $\prob(X\leq\med(X))=\frac{1}{2}\,.$

\medskip

\begin{defn}\label{perdida}
 The \textit{excess of loss} for a continuous loss random variable $X$ is the random variable: $$L\,:=\,X\,-\,\med(X).$$
\end{defn}

\medskip

As suggested by McNeil \textit{et al.}(2015) one way to interpret a \textit{risk measure} is as the required additional risk capital $\varrho(L)$ to cover a loss in excess of what it was originally estimated. In the specialized literature on this subject there are many properties for risk measures that are considered as ``desirable'' or ``reasonable'', though some concerns have been raised for some of them.

\medskip

\begin{defn}\label{monotona}
  A risk measure $\varrho$ is \textit{monotone} if for any excess of loss random variables $L_1$ and $L_2$ such that $\prob(L_1\leq L_2)=1$ we have that $\varrho(L_1)\leq\varrho(L_2).$
\end{defn}

\medskip

McNeil \textit{et al.}(2015) and several other authors consider monotonicity as a clearly desirable property since financial positions that involve higher risks under any circumstance should be covered by more risk capital. Positions such that $\varrho(L)\leq 0$ do not require additional capital.

\medskip

\begin{defn}\label{invariancia}
  A risk measure $\varrho$ is \textit{translation invariant} if for any excess of loss random variable $L$ and any constant $c$ we have that $\varrho(L+c)=\varrho(L)+c.$
\end{defn}

\medskip

This property is also considered as desirable by McNeil \textit{et al.}(2015) and other authors under the following argument: the uncertainty associated to $L':=L+c\,$ totally depends on $L$ since $c$ is fixed, $\varrho(L)$ is the additional risk capital required to cover an excess of loss under $L$ and therefore it would be enough to add the fixed amount $c$ in order to cover for $L'.$

\medskip

\begin{defn}\label{subaditividad}
  A risk measure $\varrho$ is \textit{subadditive} if for any excess of loss random variables $L_1$ and $L_2$ we have that $\varrho(L_1+L_2)\leq\varrho(L_1)+\varrho(L_2).$
\end{defn}

\medskip

This property cannot be considered as generally acceptable since there is some debate around it. One argument in favor is that diversification \textbf{always} reduces risk, which is more a \textbf{dogma} rather than something proved to be true under all circumstances. We may counter argue that for some risks there could be some sort of pernicious interaction that generates additional risk to the individual ones, so it may be also argued that it is better for a risk measure not to be subadditive, so that whenever it happens that $\varrho(L_1+L_2)>\varrho(L_1)+\varrho(L_2)$ then it becomes clear that diversification is not convenient in such case.

\medskip

\begin{defn}\label{homogeneo}
  A risk measure $\varrho$ is \textit{positively homogeneous} if for any excess of loss random variable $L$ and any constant $\lambda>0$ we have that $\varrho(\lambda L)=\lambda\varrho(L).$
\end{defn}

\medskip

With regard to this property McNeil \textit{et al.}(2015) and other authors mention that in case subadditivity has been accepted as reasonable then for any positive integer $n$ it should be accepted that
\begin{equation}\label{homosub}
  \varrho(nL)\,=\,\varrho(L\,+\,\cdots\,+\,L)\,\leq n\varrho(L)
\end{equation}
and since there is no diversification ``benefit'' (because just a single risk source is involved) then the highest value would be attained in (\ref{homosub}), that is equality. The same authors acknowledge there is some criticism about this property since for sufficiently large values of $\lambda$ we should have $\varrho(\lambda L)>\lambda\varrho(L)$ to penalize for a high concentration of risk in a single source of it.

\medskip

\begin{defn}\label{coherente}
  $\varrho$ is a  \textit{coherent risk measure} if it satisfies Definitions \ref{monotona} through \ref{homogeneo}.
\end{defn}

\medskip

The adjective ``coherent'' in this definition is somehow overbearing since it implicitly suggests that any risk measure that does not satisfy this definition would be incoherent despite the fact there is some debate and concerns about two of the four properties it requires. There are other additional properties that have been proposed in some contexts, see McNeil \textit{et al.}(2015) or Denuit \textit{et al.}(2005), but for the purpose of this article the above mentioned ones are enough.

\section{Value at risk}

As suggested by McNeil \textit{et al.}(2015) we may interpret $\varrho(L)$ as the additional risk capital to cover for a potential excess of loss with $L,$ but in practice such interpretation could be easily unachievable. Consider, for example, an insurance portfolio with certain face amounts for each issued policy. The only way to guarantee that the insurance company has enough resources to pay the claims under all possible scenarios would require the total reserve to be equal to the sum of all the face amounts in such portfolio.

\medskip

In practice, specially under the \textit{Basel Accords} and \textit{Solvency II} frameworks, what is calculated is the amount of risk capital that has an \textit{acceptable} high probability (but strictly less than $1$) of covering an excess of loss that might face an insurance or financial institution. Who determines how much is ``acceptable''? Typically the regulatory authority, but each company may decide to use probability levels even higher than the regulatory ones.

\medskip

\begin{defn}\label{VaR}
  \textit{Value at Risk} of level $0<\alpha<1$ for an excess of loss random variable $L$ is a risk measure defined as $$\text{VaR}_{\alpha}(L)\,:=\,F_L^{\,-1}(\alpha)$$ where $F_L^{\,-1}$ is the quantile function of $L,$ that is the inverse of the probability distribution function of $L.$
\end{defn}

\medskip

In other words, a level $\alpha$ Value at Risk associated to a continuous random variable is the amount that such variable would not exceed with probability $\alpha.$ It should be noticed that the median is a Value at Risk of level $\alpha=\frac{1}{2}.$

\medskip

\begin{prop}\label{VaRprops}
   VaR is a monotone, translation invariant, and positively homogeneous risk measure.
\end{prop}
\noindent\textit{Proof:}
\begin{itemize}
  \item[a)] Let $X$ and $Y$ be random variables such that $\prob(X\leq Y)=1.$ Then for any value $x\in\mathbb{R}:$ $$\prob(X\leq x)\,=\,\prob(X\leq x<Y)\,+\,\prob(X\leq Y\leq x)\,\geq\,\prob(\{X\leq Y\}\cap\{Y\leq x\})\,=\,\prob(Y\leq x)\,,$$ that is $F_X(x)\geq F_Y(x).$ Let $x_{\alpha}:=\text{VaR}_{\alpha}(X)$ and $y_{\alpha}:=\text{VaR}_{\alpha}(Y).$ Then $\alpha=F_X(x_{\alpha})\geq F_Y(x_{\alpha})$ and since $\alpha=F_Y(y_{\alpha})$ and distribution functions are non decreasing, necessarily $x_{\alpha}\leq y_{\alpha}$ and therefore $\text{VaR}_{\alpha}(X)\leq\text{VaR}_{\alpha}(Y).$
	\item[b)] Let $X$ be a continuous random variable with strictly increasing distribution function $F_X$ and let $c\in\mathbb{R}$ be any given constant. Define the random variable $Y:=X+c\,,$ its probability distribution function is: $$F_Y(y)\,=\,\prob(Y\leq y)\,=\,\prob(X+c\leq y)\,=\,\prob(X\leq y-c)\,=\,F_X(y-c).$$ Let $x_{\alpha}:=\text{VaR}_{\alpha}(X)$ and $y_{\alpha}:=\text{VaR}_{\alpha}(Y).$ Then: $$F_X(x_{\alpha})\,=\,\alpha\,=\,F_Y(y_{\alpha})\,=\,F_X(y_{\alpha}-c)\,,$$ and since $F_X$ is strictly increasing then $\,x_{\alpha}=y_{\alpha}-c\,$ which is equivalent to $\text{VaR}_{\alpha}(X)+c=\text{VaR}_{\alpha}(Y)=\text{VaR}_{\alpha}(X+c).$
	\item[c)] Let $X$ be a continuous random variable with strictly increasing distribution function $F_X$ and let $\lambda>0$ be a given constant. Define the random variable $Y:=\lambda X,$ its probability distribution function is: $$F_Y(y)\,=\,\prob(Y\leq y)\,=\,\prob(\lambda X\leq y)\,=\,\prob(X\leq y/\lambda)\,=\,F_X(y/\lambda).$$ Let $x_{\alpha}:=\text{VaR}_{\alpha}(X)$ and $y_{\alpha}:=\text{VaR}_{\alpha}(Y).$ Then: $$F_X(x_{\alpha})\,=\,\alpha\,=\,F_Y(y_{\alpha})\,=\,F_X(y_{\alpha}/\lambda)\,,$$ and since $F_X$ is strictly increasing then $\,x_{\alpha}=y_{\alpha}/\lambda\,$ which is equivalent to $\lambda\text{VaR}_{\alpha}(X)=\text{VaR}_{\alpha}(Y)=\text{VaR}_{\alpha}(\lambda X).\quad_{\pmb{\square}}$
\end{itemize}

\medskip

It should be noticed that VaR is proved to be positively homogeneous without a subadditivity argument as in (\ref{homosub}). In fact, VaR is not generally subadditive as it will become clear in a following section, but it will be also argued that this should not be considered as a disadvantage. 

\medskip

\begin{ejem}\label{pareto}
  Let $X$ be a \textit{Pareto} continuous random variable with parameters $\beta>0$ y $\delta>0.$ Its probability density function is given by: $$f_X(x\,|\,\beta,\delta)\,=\,\frac{\delta\beta^{\delta}}{x^{\delta+1}}\,,\quad x>\beta,$$ and therefore its probability distribution function: $$F_X(t)\,=\,\int_{-\infty}^{\,t}f_X(x\,|\,\beta,\delta)\,dx\,=\,\delta\beta^{\delta}\int_{\beta}^{\,t}\frac{dx}{x^{\delta+1}}\,=\,1\,-\,\bigg(\frac{\beta}{t}\bigg)^{\delta}\,,\quad t>\beta.$$ The quantile function of $X$ is the inverse of $F_X,$ that is $F_X^{\,-1}(u)=\beta(1-u)^{-1/\delta}$ for $0<u<1$ and consequently the median is $\med(X)=\text{VaR}_{1/2}(X)=F_X^{\,-1}(\frac{1}{2})=2^{1/\delta}\beta.$ The level $\alpha>\frac{1}{2}$ VaR for the excess of loss $L=X-\med(X)$ is given by: $$\text{VaR}_{\alpha}(L)\,=\,\text{VaR}_{\alpha}(X-\med(X))\,=\,\text{VaR}_{\alpha}(X)-\med(X)\,=\,\beta[(1-\alpha)^{-1/\delta}-2^{1/\delta}].$$ Thus, with probability $\alpha$ the excess of loss will not exceed the amount $\text{VaR}_{\alpha}(L).$ Notice that if $\alpha\rightarrow 1-$ then $\text{VaR}_{\alpha}(L)\rightarrow +\infty,$ which would require an infinite risk capital, something impossible in practice, and instead a value $\alpha<1$ sufficiently close to $1$ is arbitrarily set by the regulatory authority, for example $\alpha=0{.}995,$ though it is not clear how a particular value of $\alpha$ is considered ``safe enough'' in some sense.
\end{ejem}

\medskip

As an additional comment for this last example, the mean for the \textit{Pareto} model may no exist, it only does when $\delta>1$ and even in such case $\esper(X)=\beta\delta/(\delta-1)$ which implies that for values of $\delta$ sufficiently close to $1$ it is possible to have $\esper(X)>\text{VaR}_{\alpha}(X)$ for any given value $\alpha<1$ because $\lim_{\delta\rightarrow 1+}\esper(X)=+\infty.$ Since parameter $\delta$ controls tail heaviness of this probability distribution (lower $\delta$ values imply heavier right tail) this exemplifies a comment at the beginning of the previous section in the sense that it is better to use the median instead of the mean.

\section{Loss aggregation}

Consider $n$ excess of loss random variables $L_1,\ldots,L_n$ where $L_i=X_i-\med(X_i)$ for $i\in\{1,\ldots,n\}$ as in Definition \ref{perdida}. It is of interest to calculate VaR of the aggregation of such random variables:
\begin{equation}\label{agrega}
  L\,=\,L_1\,+\,\cdots\,+L_n\,=\,\sum_{i\,=\,1}^n X_i\,-\,\sum_{i\,=\,1}^n M(X_i)\,=\,S\,-\,c\,,
\end{equation}
where the random variable $S:=\sum_{i\,=\,1}^n X_i$ and the constant $c:=\sum_{i\,=\,1}^n M(X_i).$ In this case we get $\text{VaR}_{\alpha}(L)=\text{VaR}_{\alpha}(S)-c\,$ so this last calculation essentially depends on obtaining or estimating the probability distribution function of $S,$ that is $F_S,$ because $\text{VaR}_{\alpha}(S)=F_S^{\,-1}(\alpha).$ Since $S$ is a transformation of the $n$-dimensional random vector $(X_1,\ldots,X_n)$ it is necessary to know either the joint probability distribution function $F_{X_1,\ldots,X_n}(x_1,\ldots,x_n)=\prob(X_1\leq x_1,\ldots,X_n\leq x_n)$ or its joint probability density function $f_{X_1,\ldots,X_n}(x_1,\ldots,x_n)\geq 0$ such that $$\prob[\,(X_1,\ldots,X_n)\,\in\,B\,]\,=\,\int\cdots\int_{\!\!\!\!\!\!\!\!\!\!\!\!\!\!B}\;\quad f_{X_1,\ldots,X_n}(x_1,\ldots,x_n)\,dx_1\cdots dx_n\,.$$

\medskip

A very popular probabilistic model is the \textit{multivariate Normal} distribution, which undoubtedly has very nice mathematical properties that makes it very attractive for analysis and simplified calculations, but in practice it is usually inappropriate for the following reasons:
\begin{itemize}
  \item All the univariate marginal distributions have to be \textit{Normal}. Very often excess of loss random variables exhibit such a probabilistic behavior that are easily rejected by standard statistical normality tests, specially for heavier tails than the \textit{Normal} distribution. 
	\item The \textit{multivariate Normal} is completely unable to consider \textit{tail dependence} that very often is present among risks in finance and insurance, which consists in an important increase of the dependence degree under extreme values of the random variables involved.
\end{itemize}
These two flaws combined usually lead to a significant underestimation of the total aggregated risk. Instead, more flexible models have been explored, such as the ones built by means of \textit{copula functions} which allow for any kind and distinct marginal univariate distributions and also account for tail dependence. Getting into the details of copula modeling is beyond the scope of the present article, the interested reader should refer to Nelsen (2006) for a book on basic copula theory, and the books by McNeil \textit{et al.}(2015) and Denuit \textit{et al.}(2005) for applications of copulas in finance and insurance risk modeling.

\medskip

In two following sections, calculation of aggregated VaR will be considered in two extreme cases: perfect positive dependence (comonotonicity) and complete absence of dependence (that is, independence). For simplicity, but without loss of generality, it is considered the aggregation of two excess of loss random variables, that is $L=L_1+L_2$ where $L_1=X-\med(X)$ and $L_2=Y-\med(Y),$ which is equivalent to $L=S-c$ with $S:=X+Y$ and $c:=\med(X)+\med(Y)$ and therefore $\text{VaR}_{\alpha}(L)=\text{VaR}_{\alpha}(S)-c.$

\section{Comonotonicity}

The following result comes from the works by Hoeffding (1940) and Fr\'echet (1951) and it is known as the \textit{Fr\'echet-Hoeffding bounds} for joint probability distribution functions, which for simplicity is stated for the bivariate case:

\medskip

\begin{lema}\label{FH} (Fr\'echet--Hoeffding) If $(X,Y)$ is a random vector with joint probability distribution function $F_{X,Y}(x,y)=\prob(X\leq x,Y\leq y)$ and marginal distribution functions $F_X(x)=\prob(X\leq x)$ and $F_Y(y)=\prob(Y\leq y)$ then: $$H_{*}(x,y):=\max\{F_X(x)+F_Y(y)-1,0\}\,\leq\,F_{X,Y}(x,y)\,\leq\,\min\{F_X(x),F_Y(y)\}=:H^{*}(x,y)\,,$$ where the lower bound $H_{*}$ and the upper bound $H^{*}$ are both joint distribution functions and therefore infimum and supremum for all bivariate joint distribution functions.
\end{lema}

\medskip

\begin{defn}\label{comonotona}
  Two random variables $X$ and $Y$ are \textit{comonotone} or \textit{perfectly positively dependent} if there exists a strictly increasing function $g$ such that $\prob[Y=g(X)]=1.$
\end{defn}

\medskip

Proof of the following lemma may be found in Nelsen (2006) as Theorem 2.5.4 and following comment thereof:

\medskip

\begin{lema}\label{lemaNelsen} (Nelsen, 2006) Let $X$ and $Y$ be continuous random variables with marginal distribution functions $F_X$ and $F_Y,$ respectively, and joint distribution function $F_{X,Y}.$ Then $X$ and $Y$ are comonotone if and only if $F_{X,Y}$ is equal to the Fr\'echet-Hoeffding upper bound.
\end{lema}

\bigskip

Now the main result for this section:

\medskip

\begin{teo}\label{sumacomonotona} If $X$ and $Y$ are continuous comonotone random variables then: $$\text{VaR}_{\alpha}(X\,+\,Y)\,=\,\text{VaR}_{\alpha}(X)\,+\,\text{VaR}_{\alpha}(Y).$$
\end{teo}
\noindent\textit{Proof:}

\smallskip

\noindent Since $X$ and $Y$ are comonotone there exists a strictly increasing function $g$ such that $\prob[Y=g(X)]=1,$ hence the distribution function of $Y$ may be expressed as:
\begin{equation*}
  F_Y(y)\,=\,\prob(Y\leq y)\,=\,\prob[g(X)\leq y]\,=\,\prob[X\leq g^{-1}(y)]\,=\,F_X(g^{-1}(y)).
\end{equation*}
By Lemma \ref{lemaNelsen} we get:
\begin{equation*}
  F_{X,Y}(x,y) \,=\, \min\{F_X(x),F_Y(y)\}\,=\,\min\{F_X(x),F_X(g^{-1}(y))\}.
\end{equation*}
Define $S:=X+Y,$ then its distribution function satisfies:
\begin{equation*}
  F_S(s)\,=\,\prob(S\leq s)\,=\,\prob(X+Y\leq s)\,=\,\prob(X+g(X)\leq s)\,=\,\prob(Y\leq s-X).
\end{equation*}
Since $\prob[Y=g(X)]=1$ then $F_{X,Y}$ is a singular distribution because all the probability is distributed along the curve $y=g(x)$ and therefore $F_S(s)$ is equal to the value of $F_{X,Y}$ at the intersection point $(x_*,y_*)$ between the increasing curve $y=g(x)$ and the decreasing line $y=s-x,$ for all $s\in\text{Ran}\,g,$ which requires $g(x)=s-x$ and hence the intersection point is $(x_*,g(x_*))$ where $x_*$ is the solution to the equation $x+g(x)=s$ which will be denoted as $x_*=h(s).$ Since $g$ is strictly increasing so it is $h$ which has inverse $h^{-1}(x)=x+g(x).$ Then:
\begin{equation*}
  F_S(s)\,=\,F_{X,Y}(x_*,g(x_*))\,=\,\min\{F_X(x_*),F_X(g^{-1}(g(x_*)))\}\,=\,F_X(h(s)),
\end{equation*}
and consequently:
\begin{eqnarray*}
  \text{VaR}_{\alpha}(X+Y) &\,=\,& \text{VaR}_{\alpha}(S)\,=\,F_S^{\,-1}(\alpha)\,=\,h^{-1}(F_X^{\,-1}(\alpha)) \\
	                         &\,=\,& F_X^{\,-1}(\alpha)+g(F_X^{\,-1}(\alpha)) \,=\, \text{VaR}_{\alpha}(X)+\text{VaR}_{\alpha}(Y) \quad_{\pmb{\square}}
\end{eqnarray*}

\medskip

\begin{cor}\label{sumacomonotona2} If $X$ and $Y$ are continuous comonotone random variables, then for the excess of loss random variables $L_1:=X-\med(X)$ and $L_2:=Y-\med(Y)$ we have that: $$\text{VaR}_{\alpha}(L_1+L_2)\,=\,\text{VaR}_{\alpha}(L_1)\,+\,\text{VaR}_{\alpha}(L_2).$$
\end{cor}
\noindent\textit{Proof:}
\begin{eqnarray*}
  \text{VaR}_{\alpha}(L_1+L_2) &\,=\,& \text{VaR}_{\alpha}(X+Y-\med(X)-\med(Y)) \,=\, \text{VaR}_{\alpha}(X+Y)-\med(X)-\med(Y) \\
	                             &\,=\,& \text{VaR}_{\alpha}(X)-\med(X)+\text{VaR}_{\alpha}(Y)-\med(Y) \,=\, \text{VaR}_{\alpha}(L_1)+\text{VaR}_{\alpha}(L_2) \quad_{\pmb{\square}}
\end{eqnarray*}

\medskip

\begin{ejem}\label{sumaParetos}
  Let $X$ be a \textit{Pareto} random variable with parameters $\beta=1$ and $\delta>0$ and define the random variable $Y:=X^2.$ Since $Y=g(X)$ with $g(x)=x^2$ a strictly increasing function on $\text{Ran}\,X=\,]1,+\infty[\,$ then $X$ and $Y$ are comonotone, with $\text{Ran}\,Y=\,]1,+\infty[\,$ also. Making use of the formulas in Example \ref{pareto} we obtain:
	\begin{eqnarray*}
	  F_Y(y) &\,=\,& \prob(Y\leq y) \,=\, \prob(X^2\leq y) \,=\,\prob(X\leq\sqrt{y}\,) \\
		       &\,=\,& F_X(\sqrt{y}\,) \,=\, 1\,-\,\bigg(\frac{1}{y}\bigg)^{\delta/2},\quad y>1,
	\end{eqnarray*}
	which implies that $Y$ is also a \textit{Pareto} random variable but with parameters $\beta=1$ and $\delta/2,$ therefore: 
	\begin{equation*}\label{ejsumPa}
	  \text{Var}_{\alpha}(X)\,=\,(1-\alpha)^{-1/\delta}\,,\qquad \text{Var}_{\alpha}(Y)\,=\,(1-\alpha)^{-2/\delta}.
	\end{equation*}
	Now let $S:=X+Y=X+X^2$ where $\text{Ran}\,S=\,]2,+\infty[\,$ and we get:
	\begin{eqnarray*}
	    F_S(s) &\,=\,& \prob(S\leq s) \,=\, \prob(X+X^2\leq s) \,=\, \prob\big(X\leq(\sqrt{1+4s} - 1)/2\big) \\
			       &\,=\,& F_X\big((\sqrt{1+4s} - 1)/2\big) \,=\, 1 - \big(2/(\sqrt{1+4s} - 1)\big)^{\delta}\,,\quad s>2,
	\end{eqnarray*}
	from where we obtain for any $0<\alpha<1$ the following:
	\begin{equation*}
	   \text{VaR}_{\alpha}(X+Y) \,=\, \text{VaR}_{\alpha}(S) \,=\, F_S^{\,-1}(\alpha) \,=\, (1-\alpha)^{-1/\delta} + (1-\alpha)^{-2/\delta} \,=\, \text{Var}_{\alpha}(X) \,+\, \text{Var}_{\alpha}(Y)\,,
	\end{equation*}
	as expected. $\qquad_{\pmb{\square}}$
\end{ejem}

\section{Independence}

In contrast with the comonotonicity case where such property always implies that the VaR of the sum is equal to sum of the individual VaRs, under lack of dependence (independence) it is not possible to establish a general formula that relates the VaR for a sum of independent random variables to the individual VaRs, it will depend on each particular case, as it is shown in the following three examples:

\medskip

\begin{ejem}\label{VaRparetos}
  Let $X$ and $Y$ be independent and identically distributed \textit{Pareto} random variables with parameters $\beta=1$ and $\delta=1$ such that the right tail of their distributions is heavy enough for non existence of a mean. Again applying formulas from Example \ref{pareto} we get $\text{VaR}_{\alpha}(X)=(1-\alpha)^{-1}=\text{VaR}_{\alpha}(Y)$ where $0<\alpha<1$ and by independence the joint density function for the random vector $(X,Y)$ is the product of the marginal densities: $$f_{X,Y}(x,y)\,=\,f_X(x)f_Y(y)\,=\,\frac{1}{x^2 y^2}\,,\quad x>1,\,y>1.$$ Let $S:=X+Y$ then $\text{Ran}\,S=\,]2,+\infty[\,$ and its distribution function:
	\begin{eqnarray*}
	  F_S(s) &\,=\,& \prob(S\leq s) \,=\, \prob(X+Y\leq s) \,=\,\prob(Y\leq s-X) \,=\, \int\!\!\!\int_{y\,\leq\,s-x}f_{X,Y}(x,y)\,dxdy \\
		       &\,=\,& \int_1^{s-1}\!\!\!x^{-2}\!\!\int_1^{s-x}\!\!\!y^{-2}\,dydx \,=\, 1\,-\,\frac{2}{s}\,-\,\frac{2}{s^2}\log(s-1)\,,\quad s>2.
	\end{eqnarray*}
	Let $s_*:=\text{VaR}_{\alpha}(X)+\text{VaR}_{\alpha}(Y)=2/(1-\alpha)>2.$ Then: $$F_S(s_*)\,=\,\alpha\,-\,\frac{(1-\alpha)^2}{2}\log\bigg(\frac{1+\alpha}{1-\alpha}\bigg)\,<\,\alpha\,,$$ which implies that for any $0<\alpha<1:$ $$\text{VaR}_{\alpha}(X)+\text{VaR}_{\alpha}(Y)\,=\,s_*\,<\,F_S^{\,-1}(\alpha)\,=\,\text{VaR}_{\alpha}(S)\,=\,\text{VaR}_{\alpha}(X+Y).$$ Despite total absence of dependence between the random variables the right tails of their distributions are heavy enough such that the diversification effect is definitely not convenient: the VaR of the sum is greater than the sum of the individual VaRs, in this particular case. $\quad_{\pmb{\square}}$
\end{ejem}

\medskip

\begin{ejem}\label{VaRnormales}
  Now let $X$ and $Y$ be independent and identically distributed \textit{Normal}$\,(0,1)$ random variables. Their distribution function is expressed as: $$\Phi(z) \,=\,\frac{1}{\sqrt{2\pi}}\int_{-\infty}^{\,z} e^{-\,t^{\,2}/2}\,dt.$$ The tails of this distribution are not as heavy as in the previous example, and it has finite mean and variance. Then the random variable $S:=X+Y$ has \textit{Normal}$\,(0,2)$ distribution, which is the same as $\sqrt{2}\,X$ since a linear transformation of a \textit{Normal} random variable is still \textit{Normal} and $\esper(\sqrt{2}\,X)=\sqrt{2}\,\esper(X)=0$ and $\vari(\sqrt{2}\,X)=2\vari(X)=2.$ Therefore the distribution function of $S$ may be expressed as: $$F_S(s) \,=\, \prob(S\leq s) \,=\, \prob(\sqrt{2}\,X\leq s) \,=\, \prob(X\leq s/\sqrt{2})\,=\, \Phi(s/\sqrt{2})\,,$$ and its quantile function as $F_S^{\,-1}(u)=\sqrt{2}\,\Phi^{-1}(u),$ $0<u<1.$ Consequently, for any $0<\alpha<1:$ $$\text{VaR}_{\alpha}(X+Y) \,=\, \text{VaR}_{\alpha}(S) \,=\, F_S^{\,-1}(\alpha) \,=\,\sqrt{2}\,\Phi^{-1}(\alpha) \,<\, 2\Phi^{-1}(\alpha) \,=\, \text{VaR}_{\alpha}(X)+\text{VaR}_{\alpha}(Y).$$ In contrast with the previous example, the VaR of this sum of random variables is strictly less than the sum of the individual VaRs, and therefore in this particular case diversification is clearly convenient. $\quad_{\pmb{\square}}$
\end{ejem}

\medskip

\begin{ejem}\label{VaRexponenciales}
   Lastly, let $X$ and $Y$ be independent and identically distributed \textit{Exponential} random variables with parameter equal to $1.$ The right tail of this distribution is not as heavy as in Example \ref{VaRparetos} but certainly heavier than in Example \ref{VaRnormales}, with finite mean and variance. Their marginal probability density function is $f(x)=e^{-x},\,x>0,$ and the corresponding distribution function $F(x)=1-e^{-x},\,x>0,$ hence $\text{VaR}_{\alpha}(X)=-\log(1-\alpha)=\text{VaR}_{\alpha}(Y)$ where $0<\alpha<1.$ By independence the joint density function of the random vector $(X,Y)$ is the product of the marginal densities: $$f_{X,Y}(x,y)\,=\,f_X(x)f_Y(y)\,=\,e^{-(x+y)}\,,\quad x>0,\,y>0.$$ Let $S:=X+Y,$ then $\text{Ran}\,S=\,]0,+\infty[\,$ and its distribution function is:
	\begin{eqnarray*}
	  F_S(s) &\,=\,& \prob(X+Y\leq s) \,=\, \int\!\!\!\int_{y\,\leq\,s-x}f_{X,Y}(x,y)\,dxdy \\
		       &\,=\,& \int_0^s\!\!\!e^{-x}\!\!\int_0^{s-x}\!\!\!e^{-y}\,dydx \,=\,1-e^{-s}(1+s)\,,\quad s>0.
	\end{eqnarray*}
  By the way, calculating the derivative of $F_S(s)$ we get $f_S(s)=se^{-s},\,s>0,$ which is a density of a \textit{Gamma}$\,(2,1)$ random variable. Let $s_*:=\text{VaR}_{\alpha}(X)+\text{VaR}_{\alpha}(Y)=-2\log(1-\alpha).$ Then: $$g(\alpha) \,:=\,F_S(s_*)\,=\,1-(1-\alpha)^2\big(1-2\log(1-\alpha)\big)\,,\quad 0<\alpha<1.$$ By numerical approximation it is obtained that $g(\alpha)=\alpha$ if and only if $\alpha\approx 0{.}7153319,$ see Figure \ref{Ejemplo63}, $g(\alpha)<\alpha$ if $\alpha<0{.}7153319$ and $g(\alpha)>\alpha$ if $\alpha>0{.}7153319,$ which implies that
\begin{figure}[h]
  \begin{center}
    \includegraphics[width = 7.5cm, keepaspectratio]{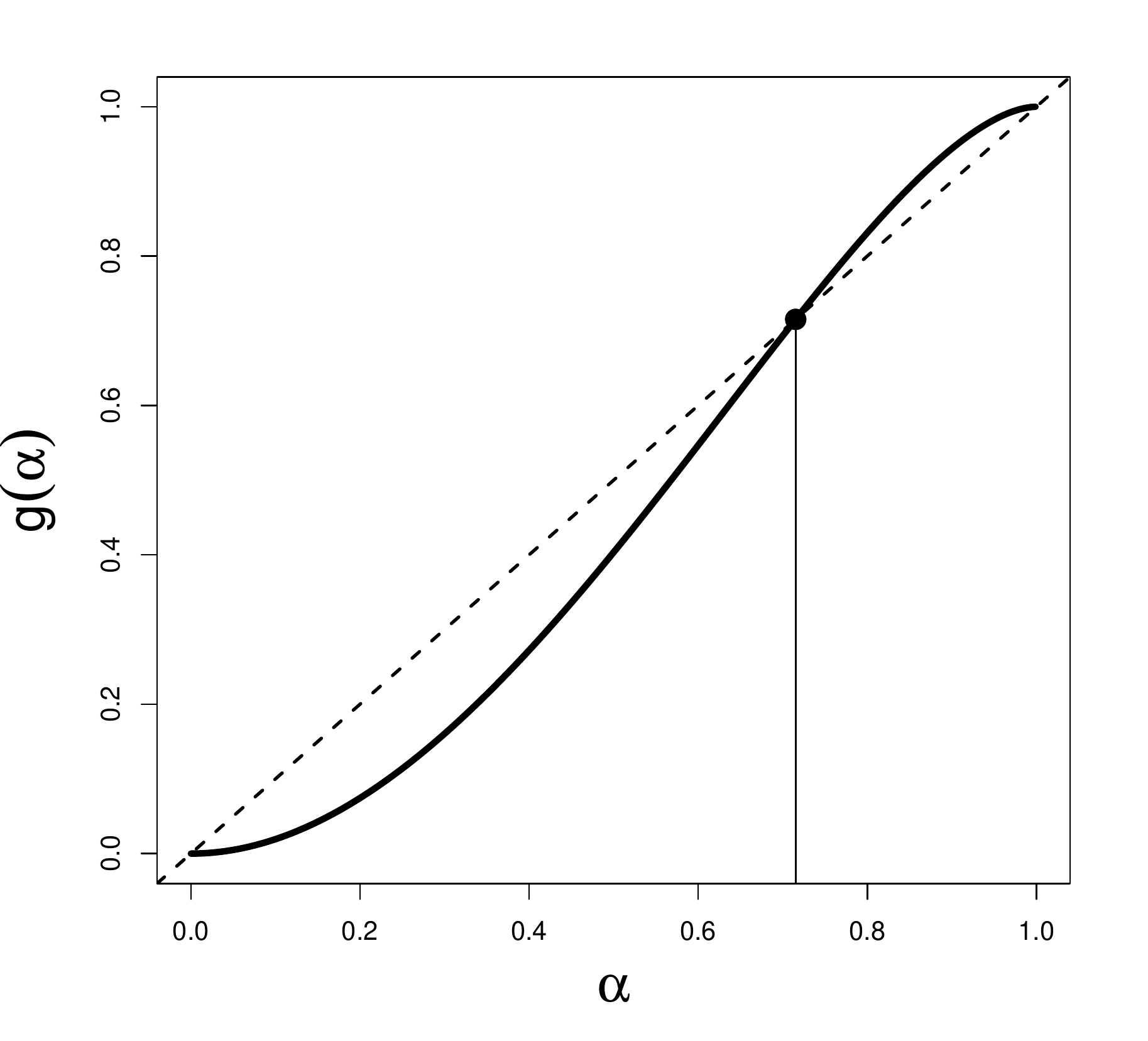}
    \caption{Graph of $g(\alpha)=F_S\big(-2\log(1-\alpha)\big)$ in Example \ref{VaRexponenciales}.}
    \label{Ejemplo63}
  \end{center}
\end{figure}
$$\text{VaR}_{\alpha}(X)+\text{VaR}_{\alpha}(Y)\,\begin{cases}
	                                                    \,<\text{VaR}_{\alpha}(X+Y) & \text{ if } \alpha<0{.}7153319 \\
																											\,=\text{VaR}_{\alpha}(X+Y) & \text{ if } \alpha\approx 0{.}7153319 \\
																											\,>\text{VaR}_{\alpha}(X+Y) & \text{ if } \alpha>0{.}7153319
	                                                 \end{cases}$$
This is an example where diversification convenience depends on the desired $\alpha$ level for VaR, in contrast with the two previous examples. $\quad_{\pmb{\square}}$
\end{ejem}

\section{Final remarks}

The main conclusion in the present work is that diversification is \textbf{not always} convenient. As shown in the examples, risk diversification may result better, worse or equivalent to lack thereof, depending on the individual risks involved and the dependence relationship between them, and even on the desired risk level. In particular, as a consequence of Theorem \ref{sumacomonotona}, if two continuous random variables are comonotone then we can guarantee that the VaR is always equal to the sum of the individual VaRs. But for independent random variables everything may happen.

\medskip

Moreover, it is argued that the fact VaR is not subadditive is more and advantage: in case the VaR of a sum is greater than the sum of individual VaRs we would be detecting a specially pernicious combination of risks on which is not convenient to diversify, while under ``coherent'' risk measures as in Definition \ref{coherente} where subadditivity is always present it would not possible to detect such a harmful risk combination.

\section*{Bibliography}

\noindent Denuit, M., Dhaene, J., Goovaerts, M., Kaas, R. (2005) \textit{Actuarial Theory for Dependent Risks.} Wiley (Chichester). 
\medskip 

%\noindent Erdely, A. (2009) C\'opulas y dependencia de variables aleatorias: una introducci\'on. \textit{Miscel\'anea matem\'atica} \textbf{48}, 7--28.
%\medskip

\noindent Fr\'echet, M. (1951) Sur les tableaux de corr\'elation dont les marges sont donn\'ees. \textit{Ann. Univ. Lyon} \textbf{14}, (Sect. A Ser. 3), 53--77.
\medskip

\noindent Hoeffding, W. (1940) Masstabinvariante Korrelationstheorie. \textit{Schriften des Matematischen Instituts und des Instituts f\"ur Angewandte Mathematik der Universit\"at Berlin} \textbf{5}, 179--223.
\medskip

\noindent McNeil, A.J., Frey, R., Embrechts, P. (2015) \textit{Quantitative Risk Management.} Princeton University Press (New Jersey).
\medskip

\noindent Nelsen, R.B. (2006) \textit{An Introduction to Copulas.} Springer (New York).
\medskip

%\noindent Sklar, A. (1959) Fonctions de r\'epartition \`a $n$ dimensions et leurs marges. \textit{Publ. Inst. Statist. Univ. Paris,} \textbf{8}, 229--231.
%\medskip

\end{document}

\maketitle

\begin{abstract}
  \noindent Se analiza el principio de diversificaci\'on de riesgos y se demuestra que no siempre resulta mejor que no diversificar, pues esto depende de caracter\'isticas individuales de los riesgos involucrados, as\'i como de la relaci\'on de dependencia entre los mismos.
\end{abstract}

\noindent \textbf{Palabras clave:} valor en riesgo, agregaci\'on de p\'erdidas, comonotonicidad, diversificaci\'on.

\section{Introducci\'on}

Un refr\'an popular sugiere que \textit{no pongas todos los huevos en una misma canasta} y lleva impl\'icito un \textit{principio} (llam\'emoslo as\'i moment\'aneamente) de \textit{diversificaci\'on de riesgos} que m\'as o menos tendr\'ia la siguiente ``justificaci\'on'': Supongamos que necesitamos trasladar $2n$ huevos caminando de un punto A a un punto B y que tenemos acceso a dos alternativas, la primera, recurrir a una persona con una sola canasta con capacidad para la totalidad de los $2n$ huevos, y la segunda, recurrir a dos personas, cada una con una canasta con capacidad para $n$ huevos, que de forma separada e independiente har\'ian dicho traslado. El mencionado refr\'an sugiere que hay mayor riesgo en la primera alternativa, pues si la persona tropieza en el camino, se romper\'ia la totalidad de los $2n$ huevos (p\'erdida total), y en cambio bajo la segunda alternativa si una de las personas tropieza s\'olo se perder\'ia la mitad, y ser\'ia muy mala suerte que ambas tropezaran, en cuyo caso la p\'erdida agregada ser\'ia de todos modos la misma que si la persona de la primera alternativa tropezara.

\medskip

Sea $X$ una variable aleatoria que cuantifica (en n\'umero de huevos) la p\'erdida bajo la primera alternativa (una sola canasta), y sea $Y$ la que cuantifica la p\'erdida bajo la segunda (dos canastas). Sea $0<\theta<1$ la probabilidad de que una persona tropiece y rompa los huevos al caminar del punto A al punto B. Entonces $X,Y$ son variables aleatorias discretas tales que $\prob(X\in\{0, 2n\})=1$ y $\prob(Y\in\{0,n,2n\})=1,$ con probabilidades puntuales $\prob(X = 2n)=\theta,$ $\prob(X = 0) = 1-\theta,$ $\prob(Y=2n)=\theta^{\hspace{0.3mm}2},$ $\prob(Y=n)=2\theta(1-\theta),$ $\prob(Y=0)=(1-\theta)^2.$ Si bien la probabilidad de tener la p\'erdida m\'axima $2n$ es mayor bajo la primera alternativa que bajo la segunda, notemos que la probabilidad de no tener p\'erdida alguna bajo la primera alternativa tambi\'en es mayor que bajo la segunda. M\'as a\'un, notemos que $$\prob(Y>0)\,=\,\theta^{\hspace{0.3mm}2}+2\theta(1-\theta)\,=\,\theta(2-\theta)\,>\,\theta\,=\,\prob(X>0)\,,$$ lo que quiere decir que es m\'as probable que suframos alg\'un tipo de p\'erdida (parcial o total) bajo la segunda alternativa que bajo la primera. Entonces... ?`es mejor poner todos los huevos en la misma canasta? Si el traslado de los huevos se va a realizar una sola vez, la respuesta ser\'ia afirmativa, pero si el mismo traslado va a realizarse un n\'umero muy grande de veces entonces deber\'iamos analizar la p\'erdida promedio de \textit{largo plazo}, que en este caso ser\'ia $\esper(X)=2n\theta$ para la primera alternativa, y $\esper(Y)=2n\theta^{\hspace{0.3mm}2}+2n\theta(1-\theta)=2n\theta$ para la segunda alternativa, es decir, a largo plazo no habr\'ia diferencia entre ambas alternativas.

\medskip

Entonces... ?`nunca conviene diversificar en dos canastas? Si la probabilidad de tropezar y romper los huevos es la misma con $2n$ huevos que con la mitad de ellos (que bien podr\'ia ser razonable hasta cierto valor de $n$) entonces el proverbio fallar\'ia, pero quiz\'as para un n\'umero $n$ suficientemente grande deber\'iamos considerar probabilidades distintas de tropezar y romperlos, digamos $\theta_1$ para la primera alternativa y $\theta_2$ para la segunda, con $\theta_1>\theta_2.$ Esta \'ultima condici\'on implicar\'ia que $\esper(X)>\esper(Y)$ y en tal caso s\'i es conveniente diversificar, siempre que se realice un n\'umero muy grande de traslados. Pero si se realiza un solo traslado entonces la condici\'on $\theta_1>\theta_2$ no ser\'ia suficiente para que convenga diversificar, a menos que $\theta_2(2-\theta_2)<\theta_1\,,$ ya que $\theta_2<\theta_2(2-\theta_2).$

\medskip

En el presente trabajo se pretende demostrar que la creencia com\'un en que diversificar riesgos \textbf{siempre} es mejor que no hacerlo, es m\'as un \textbf{dogma}\footnote{Proposici\'on tenida por cierta y como principio innegable. Conjunto de creencias de car\'acter indiscutible y obligado para los seguidores de cualquier religi\'on. \textit{Fuente:} Real Academia Espa\~nola, http://dle.rae.es/?id=E4earE8} que un principio universal cient\'ificamente comprobado, y que lo correcto es decir que diversificar riesgos puede resultar mejor, peor o igual, seg\'un el tipo de riesgos involucrados y la relaci\'on de dependencia entre ellos.

\section{Medidas de riesgo}

Sea $X$ una variable aleatoria continua, con funci\'on de distribuci\'on de probabilidades estrictamente creciente $F_X,$ que representa p\'erdida econ\'omica derivada de eventos contemplados en un contrato de seguro o inversi\'on. Para efectos pr\'acticos y sin p\'erdida de generalidad consideraremos cantidades monetarias a valor constante en el tiempo (por ejemplo, indexadas a la inflaci\'on). Como una estimaci\'on puntual de la p\'erdida puede utilizarse alguna medida de tendencia central como la media (esperanza) o la mediana, por ejemplo. Utilizaremos la mediana porque siempre existe para variables aleatorias continuas y es robusta, en contraste con la media que puede no existir o bien ser inestablemente grande bajo distribuciones de probabilidad con colas muy pesadas. La mediana se calcula por medio de la funci\'on de cuantiles (inversa de $F_X$), esto es $\med(X)=F_X^{\,-1}(\frac{1}{2})$ ya que $\prob(X\leq\med(X))=\frac{1}{2}\,.$ 

\medskip

\begin{defn}\label{perdida}
  La \textit{p\'erdida en exceso} a lo inicialmente estimado para una variable aleatoria continua $X$ que representa p\'erdidas es tambi\'en una variable aleatoria que se define: $$L\,:=\,X\,-\,\med(X).$$
\end{defn}

\medskip

Como se sugiere en McNeil \textit{et al.}(2015) una de entre varias formas para interpretar una \textit{medida de riesgo} es como la cantidad de capital adicional necesario para hacer frente a una p\'erdida en exceso que pudiera presentarse, misma que denotaremos $\varrho(L).$ Hay varias propiedades que en la literatura especializada se sugieren como ``deseables'' o ``razonables'' para cualquier medida de riesgo, algunas quiz\'as son intuitivamente razonables, otras en ocasiones generan algunos cuestionamientos.

\medskip

\begin{defn}\label{monotona}
  Una medida de riesgo $\varrho$ es \textit{mon\'otona} si para cualesquiera variables aleatorias de p\'erdida en exceso $L_1$ y $L_2$ tales que $\prob(L_1\leq L_2)=1$ se cumple $\varrho(L_1)\leq\varrho(L_2).$
\end{defn}

\medskip

Respecto a esta propiedad, McNeil \textit{et al.}(2015) y diversos autores consideran que la monotonicidad es obviamente deseable ya que posiciones o transacciones financieras que involucren mayores p\'erdidas, bajo cualquier escenario, requieren mayor capital de riesgo. Posiciones tales que $\varrho(L)\leq 0$ no requieren capital adicional alguno.

\medskip

\begin{defn}\label{invariancia}
  Una medida de riesgo $\varrho$ es \textit{invariante bajo traslaci\'on} si para cualquier variable aleatoria de p\'erdida en exceso $L$ y una constante cualquiera $c$ se cumple que $\varrho(L+c)=\varrho(L)+c.$
\end{defn}

\medskip

Tambi\'en esta propiedad es considerada como obviamente deseable por McNeil \textit{et al.}(2015) y otros autores bajo un argumento como el siguiente: la incertidumbre asociada a $L':=L+c$ depende totalmente de $L,$ no de $c,$ y por tanto si $\varrho(L)$ es el capital adicional necesario para hacer frente a una p\'erdida en exceso que pudiera presentarse con $L$ basta agregarle $c$ para contar con el capital de riesgo necesario para cubrir lo an\'alogo con $L'.$

\medskip

\begin{defn}\label{subaditividad}
  Una medida de riesgo $\varrho$ es \textit{subaditiva} si para cualesquiera variables aleatorias de p\'erdida en exceso $L_1$ y $L_2$ se cumple que $\varrho(L_1+L_2)\leq\varrho(L_1)+\varrho(L_2).$
\end{defn}

\medskip

La propiedad anterior no es considerada ``obviamente razonable'' ya que existen debates al respecto. Uno de los argumentos a favor es que diversificar \textbf{siempre} reduce el riesgo, lo cual es m\'as un \textbf{dogma} que algo que haya sido formalmente demostrado que ocurre bajo cualquier circunstancia, adem\'as de que surge la duda sobre si cierto tipo de interacci\'on entre dos o m\'as posibles fuentes de p\'erdida pudieran generar p\'erdidas adicionales a las que de por s\'i y de forma individual pueden generar. Al contrario, es opini\'on de quien escribe que es mejor que una medida de riesgo no sea subaditiva, ya que si en un momento dado ocurre que $\varrho(L_1+L_2)>\varrho(L_1)+\varrho(L_2)$ estar\'iamos detectando una combinaci\'on de riesgos especialmente perniciosa, y que por ello debi\'eramos evitar.

\medskip

\begin{defn}\label{homogeneo}
  Una medida de riesgo $\varrho$ es \textit{positivamente homog\'enea} si para cualquier variable aleatoria de p\'erdida en exceso $L$ y cualquier constante $\lambda>0$ se cumple que $\varrho(\lambda L)=\lambda\varrho(L).$
\end{defn}

\medskip

Respecto a esta propiedad McNeil \textit{et al.}(2015) y otros autores comentan que en caso de que se haya aceptado como razonable la subaditividad entonces para cualquier entero positivo $n$ se tendr\'ia que aceptar
\begin{equation}\label{homosub}
  \varrho(nL)\,=\,\varrho(L\,+\,\cdots\,+\,L)\,\leq n\varrho(L)
\end{equation}
y como no hay ``beneficio'' por diversificaci\'on al tratarse de la misma fuente de p\'erdida se alcanzar\'ia el m\'aximo valor posible en (\ref{homosub}), es decir igualdad. Los mismos autores reconocen la cr\'itica que existe respecto a esta propiedad ya que hay quienes opinan que en ciertos contextos y con valores suficientemente grandes de $\lambda$ deber\'ia cumplirse que $\varrho(\lambda L)>\lambda\varrho(L)$ para penalizar una elevada concentraci\'on del riesgo.

\medskip

\begin{defn}\label{coherente}
  $\varrho$ es una \textit{medida coherente de riesgo} si cumple con las Definiciones \ref{monotona} a \ref{homogeneo}.
\end{defn}

\medskip

El adjetivo ``coherente'' en la definici\'on anterior resulta un tanto chocante, ya que impl\'icitamente califica de incoherente a cualquier medida de riesgo que no la cumpla, a pesar de que existen cuestionamientos razonables sobre dos de las cuatro propiedades que exige. Existen propiedades adicionales que tambi\'en se han propuesto en diversos contextos, v\'ease el ya multicitado libro de McNeil \textit{et al.}(2015) o bien Denuit \textit{et al.}(2005), pero para el alcance que se pretende en el presente art\'iculo lo anterior es suficiente.

\section{Valor en riesgo}

Si bien podemos interpretar a $\varrho(L)$ como el capital de riesgo necesario para hacer frente a una p\'erdida en exceso que pudiera presentarse con $L,$ tal y como lo proponen McNeil \textit{et al.}(2015), en la pr\'actica dicha interpretaci\'on podr\'ia f\'acilmente resultar inviable. Pensemos, por ejemplo, en una cartera de p\'olizas de seguro con determinadas sumas aseguradas. La \'unica forma de garantizar que se cuenta con suficientes recursos econ\'omicos para hacer frente a todas las posibles reclamaciones es que el total de reservas de la compa\~n\'ia de seguros fuese exactamente igual a la suma de todas las sumas aseguradas en dicha cartera.

\medskip

Lo que usualmente se busca en la pr\'actica, especialmente bajo los esquemas de los \textit{Acuerdos de Basilea} y \textit{Solvencia II}\,\footnote{Cons\'ultese, por ejemplo, \textit{Un marco global para la evaluaci\'on de la solvencia del asegurador,} Informe del Grupo de Trabajo para la Evaluaci\'on de la Solvencia del Asegurador de la Asociaci\'on Actuarial Internacional (2009).}, es contar con un capital de riesgo que tenga una probabilidad ``aceptablemente alta'' (pero estrictamente menor que $1$) de cubrir el total de p\'erdida en exceso que pudiera enfrentar una entidad financiera o de seguros. ?`Qui\'en determina cu\'anto es ``aceptablemente alto''? T\'ipicamente la autoridad reguladora del sector que corresponda, aunque adicionalmente cada entidad financiera o de seguros est\'a en libertad de aplicar niveles de probabilidad a\'un mayores a los que como m\'inimo solicite el regulador.

\medskip

\begin{defn}\label{VaR}
  Se denomina \textit{valor en riesgo} de nivel $0<\alpha<1$ para una p\'erdida en exceso $L$ a la medida de riesgo denotada y definida como $$\text{VaR}_{\alpha}(L)\,:=\,F_L^{\,-1}(\alpha)$$ donde las siglas VaR corresponden en idioma ingl\'es a \textit{Value at Risk} y $F_L^{\,-1}$ es la funci\'on de cuantiles de la variable aleatoria continua $L,$ esto es, la funci\'on inversa de la funci\'on de distribuci\'on de probabilidades de $L.$
\end{defn}

\medskip

Dicho de otra forma, el valor en riesgo de nivel $\alpha$ asociado a una variable aleatoria continua es una cantidad que dicha variable no exceder\'a con probabilidad $\alpha.$ N\'otese que la mediana es un valor en riesgo de nivel $\frac{1}{2}\,.$

\medskip

\begin{prop}\label{VaRprops}
   La medida de riesgo VaR es mon\'otona, invariante bajo traslaci\'on y positivamente homog\'enea.
\end{prop}
\noindent\textit{Demostraci\'on:}
\begin{itemize}
  \item[a)] Sean $X,Y$ variables aleatorias continuas tales que $\prob(X\leq Y)=1.$ Entonces para todo valor $x\in\mathbb{R}:$ $$\prob(X\leq x)\,=\,\prob(X\leq x<Y)\,+\,\prob(X\leq Y\leq x)\,\geq\,\prob(\{X\leq Y\}\cap\{Y\leq x\})\,=\,\prob(Y\leq x)\,,$$ es decir $F_X(x)\geq F_Y(x).$ Sean $x_{\alpha}:=\text{VaR}_{\alpha}(X),$ $y_{\alpha}:=\text{VaR}_{\alpha}(Y).$ Entonces $\alpha=F_X(x_{\alpha})\geq F_Y(x_{\alpha})$ y como tambi\'en $\alpha=F_Y(y_{\alpha})$ y las funciones de distribuci\'on son mon\'otonas crecientes, necesariamente $x_{\alpha}\leq y_{\alpha}$ y por lo tanto $\text{VaR}_{\alpha}(X)\leq\text{VaR}_{\alpha}(Y).$
	\item[b)] Sea $X$ una variable aleatoria continua con funci\'on de distribuci\'on de probabilidades $F_X$ estrictamente creciente y sea $c\in\mathbb{R}$ una constante cualquiera. Definiendo una variable aleatoria $Y:=X+c$ tenemos que su funci\'on de distribuci\'on de probabilidades resulta ser: $$F_Y(y)\,=\,\prob(Y\leq y)\,=\,\prob(X+c\leq y)\,=\,\prob(X\leq y-c)\,=\,F_X(y-c).$$ Sean $x_{\alpha}:=\text{VaR}_{\alpha}(X),$ $y_{\alpha}:=\text{VaR}_{\alpha}(Y).$ Entonces: $$F_X(x_{\alpha})\,=\,\alpha\,=\,F_Y(y_{\alpha})\,=\,F_X(y_{\alpha}-c)\,,$$ y como $F_X$ es estrictamente creciente entonces necesariamente $\,x_{\alpha}=y_{\alpha}-c\,$ lo cual equivale a $\text{VaR}_{\alpha}(X)+c=\text{VaR}_{\alpha}(Y)=\text{VaR}_{\alpha}(X+c).$
	\item[c)] Sea $X$ una variable aleatoria continua con funci\'on de distribuci\'on de probabilidades $F_X$ estrictamente creciente y sea $\lambda>0$ una constante. Definiendo una variable aleatoria $Y:=\lambda X$ tenemos que su funci\'on de distribuci\'on de probabilidades resulta ser: $$F_Y(y)\,=\,\prob(Y\leq y)\,=\,\prob(\lambda X\leq y)\,=\,\prob(X\leq y/\lambda)\,=\,F_X(y/\lambda).$$ Sean $x_{\alpha}:=\text{VaR}_{\alpha}(X),$ $y_{\alpha}:=\text{VaR}_{\alpha}(Y).$ Entonces: $$F_X(x_{\alpha})\,=\,\alpha\,=\,F_Y(y_{\alpha})\,=\,F_X(y_{\alpha}/\lambda)\,,$$ y como $F_X$ es estrictamente creciente entonces necesariamente $\,x_{\alpha}=y_{\alpha}/\lambda\,$ lo cual a su vez equivale a $\lambda\text{VaR}_{\alpha}(X)=\text{VaR}_{\alpha}(Y)=\text{VaR}_{\alpha}(\lambda X).\quad_{\pmb{\square}}$
\end{itemize}

\medskip

N\'otese en particular que la medida de riesgo VaR cumple con ser positivamente homog\'enea sin necesidad de aceptar subaditividad como en (\ref{homosub}). De hecho VaR no es en general una medida de riesgo subaditiva como se mostrar\'a m\'as adelante, pero veremos tambi\'en que esto no es necesariamente una desventaja.

\medskip

\begin{ejem}\label{pareto}
  Sea $X$ una variable aleatoria continua \textit{Pareto} con par\'ametros $\beta>0$ y $\delta>0.$ Su funci\'on de densidad de probabilidades es: $$f_X(x\,|\,\beta,\delta)\,=\,\frac{\delta\beta^{\delta}}{x^{\delta+1}}\,,\quad x>\beta,$$ y por tanto su funci\'on de distribuci\'on de probabilidades resulta ser: $$F_X(t)\,=\,\int_{-\infty}^{\,t}f_X(x\,|\,\beta,\delta)\,dx\,=\,\delta\beta^{\delta}\int_{\beta}^{\,t}\frac{dx}{x^{\delta+1}}\,=\,1\,-\,\bigg(\frac{\beta}{t}\bigg)^{\delta}\,,\quad t>\beta.$$ La funci\'on de cuantiles de $X$ es la inversa de $F_X,$ esto es $F_X^{\,-1}(u)=\beta(1-u)^{-1/\delta}$ para $0<u<1$ y a partir de esto \'ultimo la mediana resulta ser $\med(X)=\text{VaR}_{1/2}(X)=F_X^{\,-1}(\frac{1}{2})=2^{1/\delta}\beta.$ Calculemos ahora el valor en riesgo de nivel $\alpha>\frac{1}{2}$ para la p\'erdida en exceso $L=X-\med(X):$ $$\text{VaR}_{\alpha}(L)\,=\,\text{VaR}_{\alpha}(X-\med(X))\,=\,\text{VaR}_{\alpha}(X)-\med(X)\,=\,\beta[(1-\alpha)^{-1/\delta}-2^{1/\delta}].$$ As\'i, con probabilidad $\alpha$ la p\'erdida en exceso no exceder\'a la cantidad $\text{VaR}_{\alpha}(L).$ N\'otese que si $\alpha\rightarrow 1-$ entonces $\text{VaR}_{\alpha}(L)\rightarrow +\infty,$ lo cual requerir\'ia un capital de riesgo infinito, algo imposible en la pr\'actica, y es por ello que normalmente se elige un valor $\alpha<1$ pero lo suficientemente cercano $1$ como para que la autoridad reguladora se sienta aceptablemente tranquila, por ejemplo $\alpha=0{.}995,$ aunque no queda claro c\'omo especificar un valor $\alpha$ que refleje un cierto nivel abstracto de ``tranquilidad''.
\end{ejem}

\medskip

Como comentario adicional respecto al ejemplo anterior, la esperanza para el modelo \textit{Pareto} no siempre existe, solo existe cuando $\delta>1$ y a\'un en tal caso se tiene que $\esper(X)=\beta\delta/(\delta-1)$ lo cual implica que con valores de $\delta$ suficientemente cercanos a $1$ por la derecha es posible que $\esper(X)$ sea mayor que $\text{VaR}_{\alpha}(X)$ para cualquier valor dado $\alpha<1$ ya que $\lim_{\delta\rightarrow 1+}\esper(X)=+\infty.$ Como el par\'ametro $\delta$ controla el grado de pesadez en la cola de la distribuci\'on de probabilidad en este modelo (a menor valor de $\delta$ mayor pesadez) esto ilustra el comentario al inicio de la secci\'on anterior en el sentido de que es m\'as conveniente utilizar la mediana en lugar de la media.

\section{Agregaci\'on de p\'erdidas}

Consideremos ahora $n$ variables aleatorias de p\'erdida en exceso $L_1,\ldots, L_n$ en donde cada una se expresa como en la Definici\'on \ref{perdida}, es decir $L_i=X_i-\med(X_i)$ para $i\in\{1,\ldots,n\}.$ Supongamos que es de inter\'es calcular el valor en riesgo para la \textit{agregaci\'on} o suma de dichas variables:
\begin{equation}\label{agrega}
  L\,=\,L_1\,+\,\cdots\,+L_n\,=\,\sum_{i\,=\,1}^n X_i\,-\,\sum_{i\,=\,1}^n M(X_i)\,=\,S\,-\,c\,,
\end{equation}
en donde se definen la variable aleatoria $S:=\sum_{i\,=\,1}^n X_i$ y la constante $c:=\sum_{i\,=\,1}^n M(X_i).$ En este caso tendr\'iamos que $\text{VaR}_{\alpha}(L)=\text{VaR}_{\alpha}(S)-c\,$ por lo que dicho c\'alculo dependende esencialmente de poder obtener o estimar la funci\'on de distribuci\'on de probabilidades de $S,$ es decir $F_S,$ ya que $\text{VaR}_{\alpha}(S)=F_S^{\,-1}(\alpha).$ Como $S$ es una transformaci\'on del vector aleatorio $n$-dimensional $(X_1,\ldots,X_n)$ entonces para la obtenci\'on o estimaci\'on de $F_S$ se requiere una distribuci\'on de probabilidades conjunta que capture adecuadamente las dependencias entre las variables aleatorias que integran dicho vector aleatorio, ya sea una funci\'on de distribuci\'on conjunta de probabilidades $F_{X_1,\ldots,X_n}(x_1,\ldots,x_n)=\prob(X_1\leq x_1,\ldots,X_n\leq x_n)$ o bien una funci\'on de densidad de probabilidades conjunta $f_{X_1,\ldots,X_n}(x_1,\ldots,x_n)\geq 0$ tal que $$\prob[\,(X_1,\ldots,X_n)\,\in\,B\,]\,=\,\int\cdots\int_{\!\!\!\!\!\!\!\!\!\!\!\!\!\!B}\;\quad f_{X_1,\ldots,X_n}(x_1,\ldots,x_n)\,dx_1\cdots dx_n\,.$$

\medskip

Un modelo probabil\'istico muy popular es la distribuci\'on de probabilidad \textit{Normal multivariada}, que si bien tiene propiedades matem\'aticas que la hacen muy atractiva para el an\'alisis y simplificar c\'alculos, resulta con mucha frecuencia un modelo inapropiado por las siguientes razones:
\begin{itemize}
  \item Las distribuciones marginales univariadas deben todas tener distribuci\'on \textit{Normal}. Con frecuencia las variables de p\'erdida exhiben distribuciones de probabilidad que son rechazadas por pruebas estad\'isticas de normalidad, t\'ipicamente por tener colas m\'as pesadas. 
	\item La distribuci\'on \textit{Normal multivariada} es incapaz de incorporar \textit{dependencia en las colas} (\textit{tail dependence} en idioma ingl\'es), una caracter\'istica que con frecuencia se observa entre variables asociadas a riesgos en seguros y finanzas y que consiste en un incremento importante en el grado de dependencia bajo valores extremos de las variables involucradas.
\end{itemize}
Estas dos deficiencias conducen a una subestimaci\'on del riesgo total agregado, lo que ha motivado la b\'usqueda de modelos probabil\'isticos m\'as flexibles, como los que se pueden construir por medio de \textit{funciones c\'opula}, mismos que permiten utilizar distribuciones marginales univariadas de cualquier tipo y distintas para cada variable involucrada, y adem\'as incorporar dependencia en las colas (tail dependence). Entrar al detalle de esto ser\'ia motivo de otro art\'iculo, por lo pronto simplemente se hace referencia a Nelsen (2006) cuyo libro es considerado fundamental para la comprensi\'on de la teor\'ia b\'asica de funciones c\'opula, y nuevamente los libros de McNeil \textit{et al.}(2015) o bien Denuit \textit{et al.}(2005) para su aplicaci\'on en finanzas y seguros. Para una breve introducci\'on  a funciones c\'opula puede revisarse Erdely (2009).

\medskip

En las dos secciones siguientes se analizar\'a el c\'alculo del valor en riesgo de una agregaci\'on de variables aleatorias en dos casos extremos: dependencia positiva perfecta (comonotonicidad) y ausencia total de dependencia (independencia). Por simplicidad pero sin p\'erdida de generalidad analizaremos el caso de la agregaci\'on de dos variables de p\'erdida en exceso, esto es $L=L_1+L_2$ donde $L_1=X-\med(X)$ y $L_2=Y-\med(Y),$ que podemos expresar como $L=S-c$ donde $S:=X+Y$ y $c:=\med(X)+\med(Y)$ y por tanto $\text{VaR}_{\alpha}(L)=\text{VaR}_{\alpha}(S)-c.$

\section{Comonotonicidad}

El siguiente es un resultado consecuencia de los trabajos de Hoeffding (1940) y Fr\'echet (1951) conocido como \textit{cotas de Fr\'echet-Hoeffding} para funciones de distribuci\'on de probabilidad conjunta, que por simplicidad aqu\'i lo restringimos al caso bivariado:

\medskip

\begin{lema}\label{FH} (Fr\'echet--Hoeffding) Si $(X,Y)$ es un vector aleatorio con funci\'on de distribuci\'on conjunta $F_{X,Y}(x,y)=\prob(X\leq x,Y\leq y)$ y funciones de distribuci\'on marginales $F_X(x)=\prob(X\leq x)$ y $F_Y(y)=\prob(Y\leq y)$ entonces: $$H_{*}(x,y):=\max\{F_X(x)+F_Y(y)-1,0\}\,\leq\,F_{X,Y}(x,y)\,\leq\,\min\{F_X(x),F_Y(y)\}=:H^{*}(x,y)\,,$$ en donde la cota inferior $H_{*}$ y la cota superior $H^{*}$ son ambas funciones de distribuci\'on conjunta y por tanto constituyen \'infimo y supremo de todas las funciones de distribuci\'on conjunta bivariadas.
\end{lema}

\medskip

\begin{defn}\label{comonotona}
  Se dice que dos variables aleatorias $X,Y$ son \textit{comon\'otonas} o bien que tienen \textit{dependencia positiva perfecta} si existe una funci\'on $g$ estrictamente creciente tal $\prob[Y=g(X)]=1.$
\end{defn}

\medskip

La demostraci\'on del siguiente lema puede consultarse en Nelsen (2006) como Teorema 2.5.4 y comentario posterior, mismo que resulta necesario para el teorema principal de esta secci\'on:

\medskip

\begin{lema}\label{lemaNelsen} (Nelsen, 2006) Sean $X,Y$ variables aleatorias continuas con funciones de distribuci\'on marginal $F_X$ y $F_Y,$ respectivamente, y funci\'on de distribuci\'on conjunta $F_{X,Y}.$ Entonces $X,Y$ son comon\'otonas si y s\'olo si $F_{X,Y}$ es igual a la cota superior de Fr\'echet-Hoeffding.
\end{lema}

\medskip

A continuaci\'on, el resultado principal de esta secci\'on:

\medskip

\begin{teo}\label{sumacomonotona} Si $X,Y$ son variables aleatorias continuas y comon\'otonas entonces: $$\text{VaR}_{\alpha}(X\,+\,Y)\,=\,\text{VaR}_{\alpha}(X)\,+\,\text{VaR}_{\alpha}(Y).$$
\end{teo}
\noindent\textit{Demostraci\'on:}

\smallskip

\noindent Como $X,Y$ son comon\'otonas entonces existe una funci\'on $g$ estrictamente creciente tal que $\prob[Y=g(X)]=1,$ por lo que la funci\'on de distribuci\'on de $Y$ puede expresarse como:
\begin{equation*}
  F_Y(y)\,=\,\prob(Y\leq y)\,=\,\prob[g(X)\leq y]\,=\,\prob[X\leq g^{-1}(y)]\,=\,F_X(g^{-1}(y)).
\end{equation*}
Aplicando el Lema \ref{lemaNelsen} tenemos que:
\begin{equation*}
  F_{X,Y}(x,y) \,=\, \min\{F_X(x),F_Y(y)\}\,=\,\min\{F_X(x),F_X(g^{-1}(y))\}.
\end{equation*}
Si se define $S:=X+Y$ entonces su funci\'on de distribuci\'on de probabilidades satisface lo siguiente:
\begin{equation*}
  F_S(s)\,=\,\prob(S\leq s)\,=\,\prob(X+Y\leq s)\,=\,\prob(X+g(X)\leq s)\,=\,\prob(Y\leq s-X).
\end{equation*}
Como $\prob[Y=g(X)]=1$ entonces $F_{X,Y}$ es una distribuci\'on singular ya que toda la probabilidad se encuentra concentrada sobre la curva $y=g(x)$ y por lo tanto $F_S(s)$ es igual al valor acumulado por $F_{X,Y}$ en el punto de intersecci\'on $(x_*,y_*)$ de la curva creciente $y=g(x)$ con la recta decreciente $y=s-x,$ para todo valor $s\in\text{Ran}\,g,$ lo cual requiere que $g(x)=s-x$ y por tanto el punto de intersecci\'on es $(x_*,g(x_*))$ donde $x_*$ es la soluci\'on de la ecuaci\'on $x+g(x)=s$ que denotaremos $x_*=h(s).$ Como $g$ es estrictamente creciente entonces $h$ tambi\'en lo es y tiene inversa $h^{-1}(x)=x+g(x).$ Entonces:
\begin{equation*}
  F_S(s)\,=\,F_{X,Y}(x_*,g(x_*))\,=\,\min\{F_X(x_*),F_X(g^{-1}(g(x_*)))\}\,=\,F_X(h(s)),
\end{equation*}
y por lo tanto:
\begin{eqnarray*}
  \text{VaR}_{\alpha}(X+Y) &\,=\,& \text{VaR}_{\alpha}(S)\,=\,F_S^{\,-1}(\alpha)\,=\,h^{-1}(F_X^{\,-1}(\alpha)) \\
	                         &\,=\,& F_X^{\,-1}(\alpha)+g(F_X^{\,-1}(\alpha)) \,=\, \text{VaR}_{\alpha}(X)+\text{VaR}_{\alpha}(Y) \quad_{\pmb{\square}}
\end{eqnarray*}

\medskip

\begin{cor}\label{sumacomonotona2} Si $X,Y$ son variables aleatorias continuas comon\'otonas que representan p\'erdidas, entonces para las variables aleatorias de exceso de p\'erdida $L_1:=X-\med(X)$ y $L_2:=Y-\med(Y)$ se cumple que: $$\text{VaR}_{\alpha}(L_1+L_2)\,=\,\text{VaR}_{\alpha}(L_1)\,+\,\text{VaR}_{\alpha}(L_2).$$
\end{cor}
\noindent\textit{Demostraci\'on:}
\begin{eqnarray*}
  \text{VaR}_{\alpha}(L_1+L_2) &\,=\,& \text{VaR}_{\alpha}(X+Y-\med(X)-\med(Y)) \,=\, \text{VaR}_{\alpha}(X+Y)-\med(X)-\med(Y) \\
	                             &\,=\,& \text{VaR}_{\alpha}(X)-\med(X)+\text{VaR}_{\alpha}(Y)-\med(Y) \,=\, \text{VaR}_{\alpha}(L_1)+\text{VaR}_{\alpha}(L_2) \quad_{\pmb{\square}}
\end{eqnarray*}

\medskip

\begin{ejem}\label{sumaParetos}
  Sea $X$ una variable aleatoria \textit{Pareto} con par\'ametros $\beta=1$ y $\delta>0$ y def\'inase la variable aleatoria $Y:=X^2.$ Como $Y=g(X)$ con $g(x)=x^2$ una funci\'on estrictamente creciente sobre $\text{Ran}\,X=\,]1,+\infty[\,$ entonces $X,Y$ son comon\'otonas, con $\text{Ran}\,Y=\,]1,+\infty[\,$ tambi\'en. Aprovechando las f\'ormulas del Ejemplo \ref{pareto} obtenemos:
	\begin{eqnarray*}
	  F_Y(y) &\,=\,& \prob(Y\leq y) \,=\, \prob(X^2\leq y) \,=\,\prob(X\leq\sqrt{y}\,) \\
		       &\,=\,& F_X(\sqrt{y}\,) \,=\, 1\,-\,\bigg(\frac{1}{y}\bigg)^{\delta/2},\quad y>1,
	\end{eqnarray*}
	lo cual implica que $Y$ tiene distribuci\'on de probabilidad \textit{Pareto} pero con par\'ametros $\beta=1$ y $\delta/2,$ y por tanto: 
	\begin{equation*}\label{ejsumPa}
	  \text{Var}_{\alpha}(X)\,=\,(1-\alpha)^{-1/\delta}\,,\qquad \text{Var}_{\alpha}(Y)\,=\,(1-\alpha)^{-2/\delta}.
	\end{equation*}
	Ahora se define la variable aleatoria $S:=X+Y=X+X^2$ donde $\text{Ran}\,S=\,]2,+\infty[\,$ y se obtiene:
	\begin{eqnarray*}
	    F_S(s) &\,=\,& \prob(S\leq s) \,=\, \prob(X+X^2\leq s) \,=\, \prob\big(X\leq(\sqrt{1+4s} - 1)/2\big) \\
			       &\,=\,& F_X\big((\sqrt{1+4s} - 1)/2\big) \,=\, 1 - \big(2/(\sqrt{1+4s} - 1)\big)^{\delta}\,,\quad s>2,
	\end{eqnarray*}
	de donde para todo valor $0<\alpha<1$ se obtiene:
	\begin{equation*}
	   \text{VaR}_{\alpha}(X+Y) \,=\, \text{VaR}_{\alpha}(S) \,=\, F_S^{\,-1}(\alpha) \,=\, (1-\alpha)^{-1/\delta} + (1-\alpha)^{-2/\delta} \,=\, \text{Var}_{\alpha}(X) \,+\, \text{Var}_{\alpha}(Y)\,,
	\end{equation*}
	como era de esperarse. $\qquad_{\pmb{\square}}$
\end{ejem}

\section{Independencia}

En contraste con el caso de comonotonicidad donde dicha caracter\'istica implica necesariamente que el valor en riesgo de una suma de variables aleatorias es igual a la suma de los valores en riesgo individuales, bajo ausencia total de dependencia (es decir, independencia) no es posible establecer de forma general qu\'e tipo de relaci\'on existir\'a entre el valor en riesgo de dicha suma y la suma de los valores en riesgo individuales, depender\'a de cada caso particular, y para demostrarlo bastar\'a con los siguientes tres ejemplos:

\medskip

\begin{ejem}\label{VaRparetos}
  Sean $X,Y$ variables aleatorias independientes e id\'enticamente distribuidas \textit{Pareto} con par\'ametros $\beta=1$ y $\delta=1,$ que con dichos valores de los par\'ametros la cola derecha de la distribuci\'on de probabilidad es suficientemente pesada como para que esperanza y varianza no existan. Aprovechando las f\'ormulas del Ejemplo \ref{pareto} obtenemos que $\text{VaR}_{\alpha}(X)=(1-\alpha)^{-1}=\text{VaR}_{\alpha}(Y)$ donde $0<\alpha<1$ y adem\'as, por independencia, la funci\'on de densidad conjunta del vector aleatorio $(X,Y)$ es igual al producto de las densidades marginales, esto es: $$f_{X,Y}(x,y)\,=\,f_X(x)f_Y(y)\,=\,\frac{1}{x^2 y^2}\,,\quad x>1,\,y>1.$$ Si se define la variable aleatoria $S:=X+Y$ entonces $\text{Ran}\,S=\,]2,+\infty[\,$ y su funci\'on de distribuci\'on de probabilidades es:
	\begin{eqnarray*}
	  F_S(s) &\,=\,& \prob(S\leq s) \,=\, \prob(X+Y\leq s) \,=\,\prob(Y\leq s-X) \,=\, \int\!\!\!\int_{y\,\leq\,s-x}f_{X,Y}(x,y)\,dxdy \\
		       &\,=\,& \int_1^{s-1}\!\!\!x^{-2}\!\!\int_1^{s-x}\!\!\!y^{-2}\,dydx \,=\, 1\,-\,\frac{2}{s}\,-\,\frac{2}{s^2}\log(s-1)\,,\quad s>2.
	\end{eqnarray*}
	Sea $s_*:=\text{VaR}_{\alpha}(X)+\text{VaR}_{\alpha}(Y)=2/(1-\alpha)>2.$ Entonces: $$F_S(s_*)\,=\,\alpha\,-\,\frac{(1-\alpha)^2}{2}\log\bigg(\frac{1+\alpha}{1-\alpha}\bigg)\,<\,\alpha\,,$$ lo cual implica que para todo valor $0<\alpha<1:$ $$\text{VaR}_{\alpha}(X)+\text{VaR}_{\alpha}(Y)\,=\,s_*\,<\,F_S^{\,-1}(\alpha)\,=\,\text{VaR}_{\alpha}(S)\,=\,\text{VaR}_{\alpha}(X+Y).$$ A pesar de la ausencia de dependencia alguna entre las variables aleatorias involucradas, las colas de sus distribuciones individuales son suficientemente pesadas como para que el efecto de diversificaci\'on resulte inconveniente, pues el valor en riesgo de su suma resulta mayor que la suma de los valores en riesgo individuales. $\quad_{\pmb{\square}}$
\end{ejem}

\medskip

\begin{ejem}\label{VaRnormales}
  Ahora sean $X,Y$ variables aleatorias independientes e id\'enticamente distribuidas \textit{Normal}$\,(0,1)$ cuya funci\'on de distribuci\'on de probabilidades se obtiene mediante: $$\Phi(z) \,=\,\frac{1}{\sqrt{2\pi}}\int_{-\infty}^{\,z} e^{-\,t^{\,2}/2}\,dt.$$ Las colas de esta distribuci\'on de probabilidades no son tan pesadas como las del ejemplo anterior y tiene esperanza y varianza finitas. Es un conocido y elemental resultado de probabilidad que la variable aleatoria $S:=X+Y$ tiene distribuci\'on de probabilidad \textit{Normal}$\,(0,2)$ y por ello $S$ tiene la misma distribuci\'on de probabilidad que $\sqrt{2}\,X$ ya que cualquier transformaci\'on lineal de una variable aleatoria \textit{Normal} sigue siendo \textit{Normal} y adem\'as $\esper(\sqrt{2}\,X)=\sqrt{2}\,\esper(X)=0$ y $\vari(\sqrt{2}\,X)=2\vari(X)=2.$ Por lo anterior, la funci\'on de distribuci\'on de probabilidades de $S$ puede expresarse de la siguiente manera: $$F_S(s) \,=\, \prob(S\leq s) \,=\, \prob(\sqrt{2}\,X\leq s) \,=\, \prob(X\leq s/\sqrt{2})\,=\, \Phi(s/\sqrt{2})\,,$$ y su funci\'on de cuantiles mediante $F_S^{\,-1}(u)=\sqrt{2}\,\Phi^{-1}(u),$ $0<u<1,$ por lo que para todo valor $0<\alpha<1:$ $$\text{VaR}_{\alpha}(X+Y) \,=\, \text{VaR}_{\alpha}(S) \,=\, F_S^{\,-1}(\alpha) \,=\,\sqrt{2}\,\Phi^{-1}(\alpha) \,<\, 2\Phi^{-1}(\alpha) \,=\, \text{VaR}_{\alpha}(X)+\text{VaR}_{\alpha}(Y).$$ En contraste con el ejemplo anterior, el valor en riesgo de esta suma de variables aleatorias independientes es menor que la suma de los valores en riesgo individuales, y por tanto en este caso particular la diversificaci\'on resulta conveniente. $\quad_{\pmb{\square}}$
\end{ejem}

\medskip

\begin{ejem}\label{VaRexponenciales}
   Finalmente sean $X,Y$ variables aleatorias independientes e id\'enticamente distribuidas \textit{Exponencial} est\'andar (par\'ametro igual a $1$). La cola derecha de esta distribuci\'on de probabilidad no es tan pesada como la del Ejemplo \ref{VaRparetos} pero s\'i m\'as pesada que en el Ejemplo \ref{VaRnormales}, y tiene esperanza y varianza finitas. Su funci\'on de densidad de probabilidades es $f(x)=e^{-x},\,x>0,$ su funci\'on de distribuci\'on $F(x)=1-e^{-x},\,x>0$ y $\text{VaR}_{\alpha}(X)=-\log(1-\alpha)=\text{VaR}_{\alpha}(Y)$ donde $0<\alpha<1.$ Adem\'as, por independencia, la funci\'on de densidad conjunta del vector aleatorio $(X,Y)$ es igual al producto de las densidades marginales, esto es: $$f_{X,Y}(x,y)\,=\,f_X(x)f_Y(y)\,=\,e^{-(x+y)}\,,\quad x>0,\,y>0.$$ Si se define la variable aleatoria $S:=X+Y$ entonces $\text{Ran}\,S=\,]0,+\infty[\,$ y su funci\'on de distribuci\'on de probabilidades es:
	\begin{eqnarray*}
	  F_S(s) &\,=\,& \prob(X+Y\leq s) \,=\, \int\!\!\!\int_{y\,\leq\,s-x}f_{X,Y}(x,y)\,dxdy \\
		       &\,=\,& \int_0^s\!\!\!e^{-x}\!\!\int_0^{s-x}\!\!\!e^{-y}\,dydx \,=\,1-e^{-s}(1+s)\,,\quad s>0.
	\end{eqnarray*}
  Y de hecho derivando $F_S(s)$ la funci\'on de densidad resultante es $f_S(s)=se^{-s},\,s>0,$ que corresponde a una distribuci\'on de probabilidad \textit{Gamma}$\,(2,1).$ Sea $s_*:=\text{VaR}_{\alpha}(X)+\text{VaR}_{\alpha}(Y)=-2\log(1-\alpha).$ Entonces: $$g(\alpha) \,:=\,F_S(s_*)\,=\,1-(1-\alpha)^2\big(1-2\log(1-\alpha)\big)\,,\quad 0<\alpha<1.$$ Por aproximaci\'on num\'erica es verificable que $g(\alpha)=\alpha$ si y s\'olo si $\alpha\approx 0{.}7153319,$ ver Figura \ref{Ejemplo63}, que $g(\alpha)<\alpha$ si $\alpha<0{.}7153319$ y $g(\alpha)>\alpha$ si $\alpha>0{.}7153319,$ lo cual implica que
\begin{figure}[h]
  \begin{center}
    \includegraphics[width = 7.5cm, keepaspectratio]{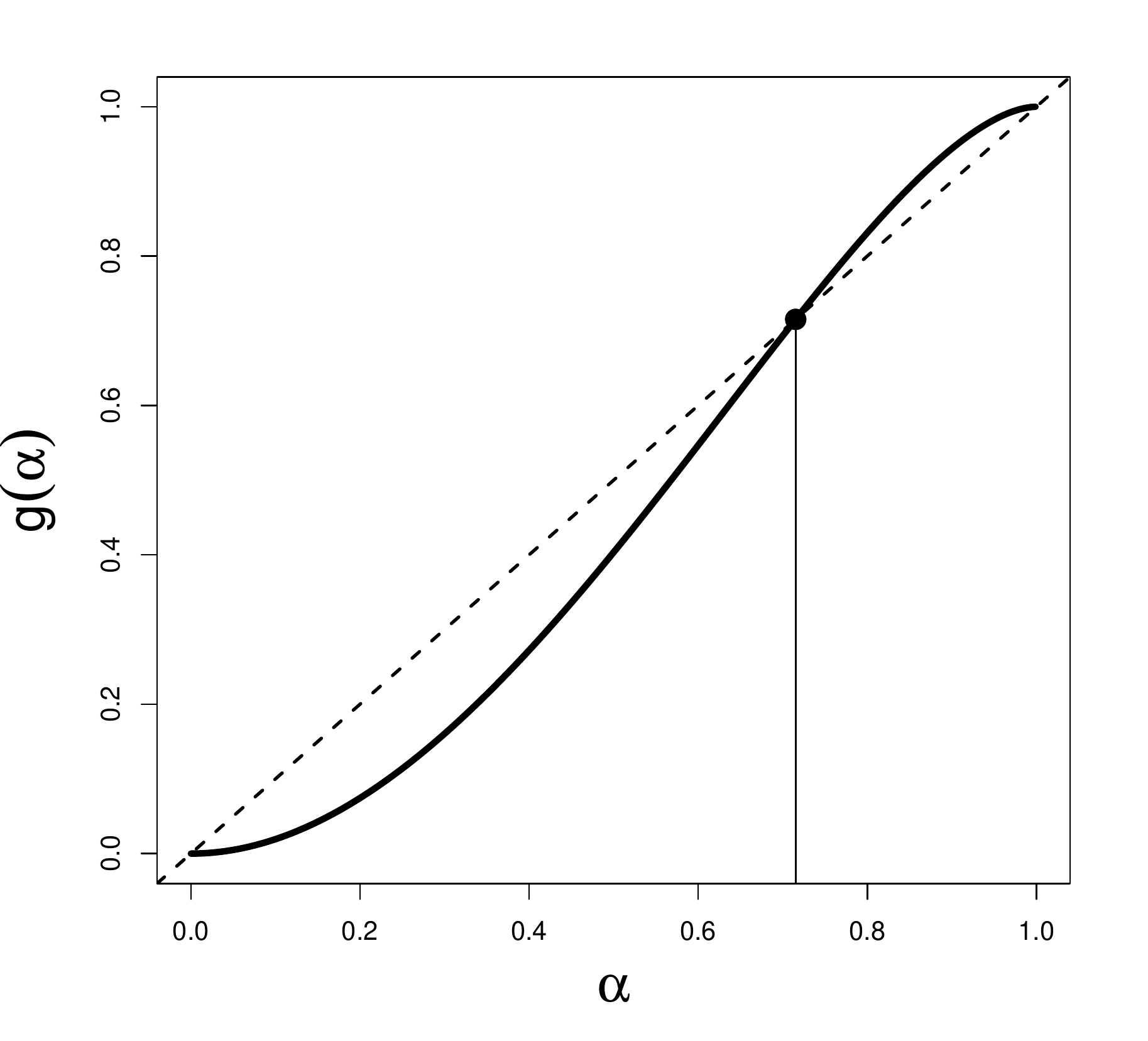}
    \caption{Gr\'afica de $g(\alpha)=F_S\big(-2\log(1-\alpha)\big)$ en el Ejemplo \ref{VaRexponenciales}.}
    \label{Ejemplo63}
  \end{center}
\end{figure}
$$\text{VaR}_{\alpha}(X)+\text{VaR}_{\alpha}(Y)\,\begin{cases}
	                                                    \,<\text{VaR}_{\alpha}(X+Y) & \text{ si } \alpha<0{.}7153319 \\
																											\,=\text{VaR}_{\alpha}(X+Y) & \text{ si } \alpha\approx 0{.}7153319 \\
																											\,>\text{VaR}_{\alpha}(X+Y) & \text{ si } \alpha>0{.}7153319
	                                                 \end{cases}$$
Aqu\'i estamos ante un ejemplo en que la conveniencia o no de diversificar depende del nivel $\alpha$ de valor en riesgo deseado, situaci\'on que no ocurri\'o en los dos ejemplos anteriores. $\quad_{\pmb{\square}}$
\end{ejem}

\section{Conclusiones}

La principal conclusi\'on del presente an\'alisis es que resulta falsa la idea de que diversificar riesgos siempre es mejor que no hacerlo. Como se pudo ilustrar en diversos ejemplos, diversificar riesgos puede resultar mejor, peor o igual que no hacerlo, dependiendo de las caracter\'isticas individuales de los riesgos involucrados, de la relaci\'on de dependencia entre ellos, e incluso hasta del nivel de riesgo deseado. En particular, como consecuencia del Teorema \ref{sumacomonotona}, si dos variables aleatorias continuas son comon\'otonas entonces s\'i es posible afirmar que en general el valor en riesgo de la suma es igual a la suma de los valores en riesgo individuales. Pero en el caso de variables aleatorias independendientes puede resultar mejor, peor o igual diversificar, dependiendo de las distribuciones de probabilidad de los riesgos involucrados.

\medskip

Por lo anterior, se argumenta tambi\'en que la no subaditividad del VaR es m\'as una ventaja que desventaja, pues en aquellos casos en que el VaR de una agregaci\'on de riesgos resulte mayor que la suma de los VaR individuales estar\'iamos obteniendo inmediatamente informaci\'on sobre cierto tipo de interacci\'on entre riesgos que conviene evitar. En cambio, con medidas ``coherentes'' de riesgo como en la Definici\'on \ref{coherente} donde la subaditividad est\'a garantizada, no tendr\'iamos aviso alguno sobre interacciones de riesgo perniciosas.

\section*{Bibliograf\'ia}

\noindent Denuit, M., Dhaene, J., Goovaerts, M., Kaas, R. (2005) \textit{Actuarial Theory for Dependent Risks.} Wiley (Chichester). 
\medskip 

\noindent Erdely, A. (2009) C\'opulas y dependencia de variables aleatorias: una introducci\'on. \textit{Miscel\'anea matem\'atica} \textbf{48}, 7--28.
\medskip

\noindent Fr\'echet, M. (1951) Sur les tableaux de corr\'elation dont les marges sont donn\'ees. \textit{Ann. Univ. Lyon} \textbf{14}, (Sect. A Ser. 3), 53--77.
\medskip

\noindent Hoeffding, W. (1940) Masstabinvariante Korrelationstheorie. \textit{Schriften des Matematischen Instituts und des Instituts f\"ur Angewandte Mathematik der Universit\"at Berlin} \textbf{5}, 179--223.
\medskip

\noindent McNeil, A.J., Frey, R., Embrechts, P. (2015) \textit{Quantitative Risk Management.} Princeton University Press (New Jersey).
\medskip

\noindent Nelsen, R.B. (2006) \textit{An Introduction to Copulas.} Springer (New York).
\medskip

%\noindent Sklar, A. (1959) Fonctions de r\'epartition \`a $n$ dimensions et leurs marges. \textit{Publ. Inst. Statist. Univ. Paris,} \textbf{8}, 229--231.
%\medskip